\documentclass[aps,pra,superscriptaddress,notitlepage,onecolumn, 11pt]{revtex4-2}
\usepackage{hyperref}
\usepackage{amsmath}
\usepackage{graphicx}
\usepackage{color}
\usepackage{braket}
\usepackage{epstopdf} 
\usepackage{physics}

\newcommand{\be}{\begin{equation}}
\newcommand{\ee}{\end{equation}}

\usepackage{bm}

\newcommand{\addcqt}{Centre for Quantum Technologies, National University of Singapore, 3 Science Drive 2, Singapore 117543}
\newcommand{\addtuc}{School of Electrical and Computer Engineering, Technical University of Crete, Chania, Greece 73100}
\newcommand{\addangelq}{AngelQ Quantum Computing, 531A, Upper Cross Street, \#04-95 Hong Lim Complex, Singapore 051531}

\begin{document}

\title{Unsupervised learning of quantum many-body scars using intrinsic dimension}

\author{Harvey Cao}
\affiliation{\addcqt}
\author{Dimitris G. Angelakis}
\affiliation{\addcqt}
\affiliation{\addtuc}
\affiliation{\addangelq}
\author{Daniel Leykam}
\affiliation{\addcqt}

\date{\today}

\begin{abstract}
Quantum many-body scarred systems contain both thermal and non-thermal scar eigenstates in their spectra. When these systems are quenched from special initial states which share high overlap with scar eigenstates, the system undergoes dynamics with atypically slow relaxation and periodic revival. This scarring phenomenon poses a potential avenue for circumventing decoherence in various quantum engineering applications. Given access to an unknown scar system, current approaches for identification of special states leading to non-thermal dynamics rely on costly measures such as entanglement entropy. In this work, we show how two dimensionality reduction techniques, multidimensional scaling and intrinsic dimension estimation, can be used to learn structural properties of dynamics in the PXP model and distinguish between thermal and scar initial states. The latter method is shown to be robust against limited sample sizes and experimental measurement errors.
\end{abstract}

\maketitle

\section{Introduction}
Quantum many-body scars (QMBS) are an interesting and recent paradigm describing the weak violation of ergodicity that arises from an intriguing interplay between integrability and chaos in quantum many-body systems. The predictions made by the Eigenstate Thermalization Hypothesis (ETH) no longer remain valid across the full spectrum \cite{Deutsch_2018}; a subset of atypical, non-thermal eigenstates exhibit slow relaxation dynamics which retain memory of the system's initial configuration. In particular, these QMBS eigenstates have been found to share high overlap with certain non-equilibrium states which results in persistent oscillations of local observables when the system is quenched from such a state, whereas for generic states the system quickly approaches thermal equilibrium and effectively `forgets' its initial condition \cite{Turner_2018}.

The set of scar states is exponentially small with respect to the Hilbert space dimension, which raises the question as to whether this phenomenon is too rare to be interesting or experimentally relevant. However, the discovery of anomalously long-lived revivals in a Rydberg blockaded atom array setup showed that not only can QMBS signatures be captured in experiment, but their high overlap with short-range correlated product states makes them surprisingly accessible \cite{Bernien2017}. Following this initial experimental observation, a lively search for other models hosting scar eigenstates has arisen, as well as analytical construction through direct embedding of scars into the spectrum. This has simultaneously driven work on developing a coherent theory to explain the ergodicity-breaking phenomenon \cite{Schecter_2019, Chandran_2023,Lee_2020, Moudgalya_2018, Mark_2020, PhysRevB.107.L201105, PhysRevB.107.205112}.

Whilst rapid experimental developments in quantum simulation platforms with low levels of environmental decoherence and high levels of controllability have enabled the exploration of non-equilibrium dynamics of QMBS systems, there are yet to emerge robust analytical methods for identifying systems that belong to non-analytical scarring regimes. The current procedures rely heavily on studying directly measures of return probability or entanglement entropy in order to distinguish them \cite{Schecter_2019, Iadecola_2020, Scherg_2021, Hudomal_2022, Su_2023}. These are not very experimentally accessible, especially as we scale to larger system sizes. Given limited a priori knowledge of the system model, it is important to find a scheme which is not reliant on exponentially many measurements for calculating expectation values or building a full tomographic picture in order to identify the states that exhibit special non-thermal characteristics.

\begin{figure}
  \centering
  {\includegraphics[width=0.8\textwidth]{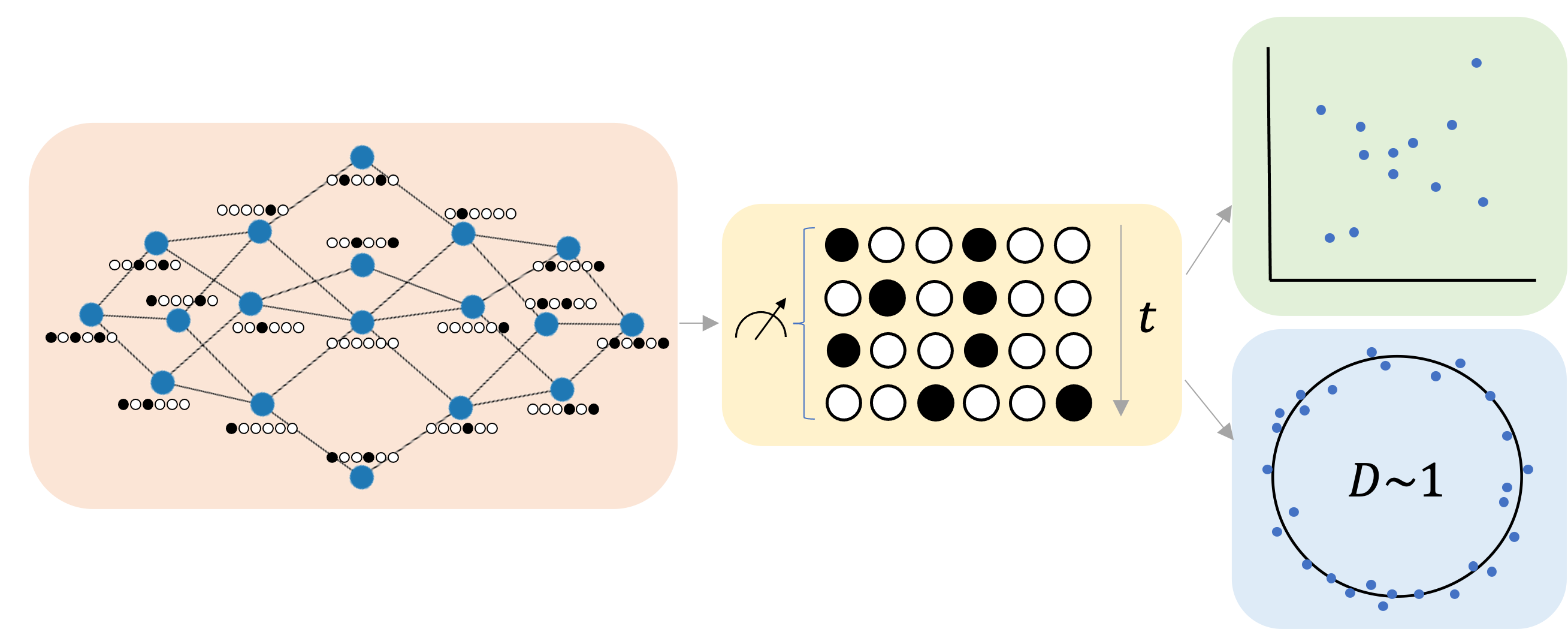}}%
  \caption{Graphical illustration of pipeline for studying quantum scar models using machine learning techniques. Hilbert space graph for PXP model (L=4) with links between spin configurations where there is an allowed transition (red). The system is sampled $P$ times at $N$ different timesteps (yellow). Multidimensional scaling embeds this data in a lower-dimension while preserving distances in a chosen metric (green) and intrinsic dimension estimation extracts the underlying structure directly from the sampled data (blue).}
  \label{fig:fig1}
\end{figure}

Tools from the field of machine learning have proven to be astonishingly effective at tackling problems that involve the analysis of complex, high-dimensional datasets. In particular, these techniques have found fruitful applications in the context of quantum many-body physics and statistical physics~\cite{RevModPhys.91.045002,doi:10.1080/23746149.2020.1797528}. The outstanding performance of these methods have often been attributed to the idea that this high dimensionality comes about only from the embedding space, while the data itself lies in a manifold with significantly fewer meaningful degrees of freedom. As a result, datasets that might be initially hard to describe can have their descriptions condensed into several key features of the latent manifold which are easier to learn. For example, a few recent works apply dimensionality reduction techniques to raw spin configuration data of spin models to identify distinct phases, using this subsequently to extract salient features such as order parameter and structure factor \cite{Wang_2016, Hu_2017, Lidiak_2020, Rodriguez-Nieva_2019,doi:10.1080/23746149.2023.2202331}. 

In this work, we consider two ML techniques in the context of manifold learning to study the scar properties of QMBS systems. The first of these is multidimensional scaling (MDS), a nonlinear dimensionality reduction method which allows high dimensional data to be translated into a lower dimension whilst approximately preserving a measure of distance between data points. Observing that there is dynamical structure in the lower-dimensional manifold, we then investigate the property of intrinsic dimension (ID), which describes the effective number of variables required for minimal representation of a dataset. Through applying an estimation technique to extract the ID directly from shot measurements of the system, we find that the manifold for the PXP model can be distinguished from thermal states in an unsupervised manner. In particular, we show that this scheme is robust to limited shots and realistic measurement noise, making it feasible for experimental implementation, e.g. using Rydberg atom quantum simulators~\cite{doi:10.1126/science.abg2530}. The graphical illustrations in Fig. \ref{fig:fig1} demonstrate the pipeline for our approach, which uses sampled data from the PXP model to both embed in a low-dimensional space and obtain an intrinsic dimension estimate. We note that there has been some recent interest in the detection of quantum scars using neural network approaches \cite{Szoldra_2022, Han_2023}. These, however, are not truly unsupervised methods and require partial training of the neural networks using existing knowledge of some scar states, which may not always be accessible. They are also limited by their scalibility to larger system sizes, which further motivates the use of our non-parametric ML models.

The outline of this paper is as follows: Section II provides a brief introduction to quantum many-body scarring and the PXP model. Section III presents a dimensionality reduction scheme for analysing dynamical many-body states using multidimensional scaling and a physics-informed distance measure. Section IV proposes an unsupervised approach for identifying many-body scar and thermalising initial states using an intrinsic dimension estimation method. Section V discusses the results obtained using these approaches, makes concluding remarks and suggests future directions.

\section{Quantum many-body scar model}

\begin{figure}
  \centering
{\includegraphics[width=0.9\textwidth]{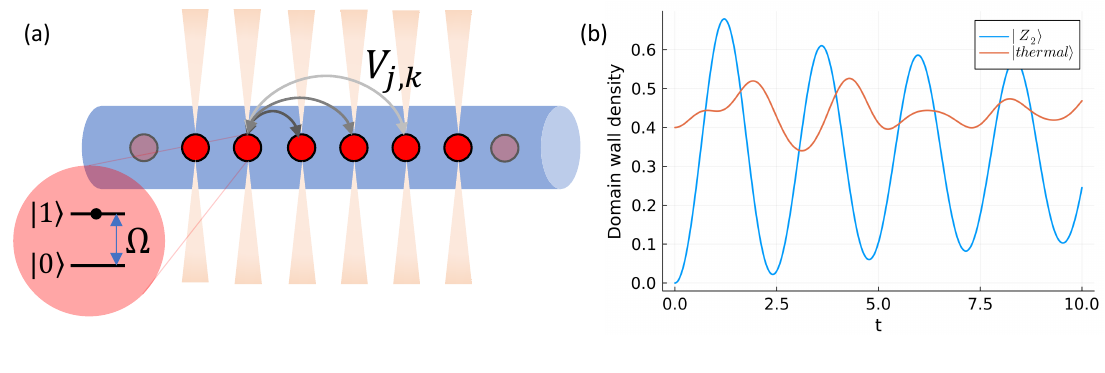}}%
  \caption{PXP system and dynamics: (a) Neutral atoms are trapped in an array using optical tweezers and driven with a laser of Rabi frequency $\Omega$, enabling transitions to a Rydberg state and van der Waals interactions $V_{j,k}$ between the atoms. (b) Persistent revivals in the domain wall density are observed when the PXP system is quenched from the $\ket{Z_{2}}$ initial state (blue line), but not from other thermalising initial states (orange line). Domain walls occur when two neighbouring atoms are in the same state or a ground state atom is at the edge of the array, and the density is computed by considering all neighbouring atom pairs across the entire chain.}
  \label{fig:fig2}
\end{figure}

In an array of Rydberg atoms, the individual atoms may be viewed as an effective two level system with atoms occupying either a ground or Rydberg excited state, which we denote by $\ket{0}$ and $\ket{1}$, respectively. When subject to a driving laser field, the atoms undergo local Rabi oscillations and are free to flip between the two states, as illustrated in Fig. \ref{fig:fig2}(a). The Hamiltonian describing a Rydberg chain of length $N$ is given by \cite{Bernien2017}
\begin{equation}
    H_{\mathrm{Rydberg}} = \sum_{i}^{N} \left(\frac{\Omega}{2}X_{i} - \Delta n_{i}\right) + \sum_{j<k}^{N}V_{j,k}n_{j}n_{k},
\end{equation}
where the parentheses include a Rabi precession term represented by the Pauli $X_{i} = \ketbra{0}{1}+\ketbra{1}{0}$ operator acting on site $i$ driven at frequency $\Omega$ and a local density term $n_{i} = \ket{1_{i}}\bra{1_{i}}$ with laser detuning $\Delta$. If the Rydberg atoms are assembled close together, atoms in excited states will interact via repulsive van der Waals forces, with an interaction strength $V_{j,k} \propto 1/R^6$, where $R$ is the distance between the atoms $j$ and $k$. In the limit where the atoms are brought very close together and the strength of nearest neighbour interactions is tuned to be much larger in amplitude than the detuning and Rabi frequency of the laser field ($\Delta=0, V\gg\Omega$), a regime of the so-called Rydberg blockade can be achieved where simultaneous excitations of neighbouring Rydberg atoms are energetically prohibited \cite{Turner_2018}. Correspondingly, the system and its Hilbert space becomes kinetically constrained in a way that forbids configurations of the form $\ket{\cdot\cdot 1 1 \cdot \cdot}$. This model is referred to as the PXP model for the form of its Hamiltonian, expressed as
\begin{equation}
    H_{PXP} = \Omega\sum_{i} P_{i-1}X_{i}P_{i+1}.
\end{equation} 
The excitation constraints are enforced by the projectors $P_{j}=\ketbra{0_{j}}{0_{j}}$, which allows a site $j$ to enter the Rydberg excited state only if both of its neighbours are in the atomic ground state. For the rest of this work, we use periodic boundary conditions (PBC) and normalize time $t$ by the Rabi frequency which we set to $\Omega=1$.

This constrained model was shown in simulations~\cite{Turner_2018} and experiments~\cite{Bernien2017} to exhibit a peculiar absence of thermalization when the system was quenched from certain special initial states such as the $\ket{Z_{2}}=\ket{01\cdot \cdot\cdot01}$ (Néel) state. This was demonstrated through the observation of periodic revival behaviour in local observables such as density of domain walls (see Fig. \ref{fig:fig2}(b)), which occur where two neighbouring atoms are in the same state or a ground state atom is at the edge of the array. In contrast, fast relaxation to a thermal value is seen when quenched from other initial states such as $\ket{\bm{0}} = \ket{0 \cdot \cdot \cdot 0}$, and remains near that value for future times. The slow relaxation phenomenon of local observables arises directly from the ability of the system to exhibit quantum wavefunction revival when quenched from the special states. The origin of this behaviour can be understood intuitively by examining the dynamics of the PXP model on a Hilbert space graph, as illustrated in Fig. \ref{fig:fig1}. The dynamics is sufficiently constrained by the structure of the graph such that the wavefunction finds itself returning to its initial point through serendipitous propagation and reflection when initialized from certain nodes.

Without knowledge of the entire spectrum, there is no universal framework for determining the special initial states that allow these non-thermalization properties to be observed in the system dynamics. Significant effort has been directed to the discovery of novel QMBS models, typically with the scar states embedded directly into the spectrum \cite{Omiya_2023,Wang_2023}. However, for kinetically constrained models like the PXP model where the scarring properties are analytically unknown, it is a difficult task to detect where the scar manifold lies. In this work, we employ machine learning techniques to learn properties of the scar manifold and develop an unsupervised scheme for detecting the special initial states of an unknown QMBS model.

\section{Dimensionality reduction of QMBS}

In recent years, the daunting task of navigating the high dimensional Hilbert spaces of quantum many-body systems has been tackled by techniques from machine learning. Their effectiveness in this application stems from the idea that the underlying manifolds of most natural datasets are smaller in dimension than the raw embedded dataset, known as the manifold hypothesis. Thus, dimensionality reduction techniques can be used to convert the unwieldy, high-dimensional data found in quantum many-body systems into a more compact form while still preserving its key features. These include methods such as principal component analysis (PCA) and diffusion maps, which are used for clustering or visualisation by projecting the data to a lower-dimensional space through minimization of an objective error function \cite{Wang_2016, Lidiak_2020}.

Multidimensional scaling (MDS) is another nonlinear optimization-based method for manifold learning that seeks to embed high-dimensional data into a lower-dimensional space. This mapping is found while preserving the pairwise distances between data points as much as possible, where the metric defining the distances can be chosen by the user. If the Euclidean distance metric is used, this method becomes equivalent to PCA, as minimizing linear distance becomes equivalent to maximising linear correlations. MDS is generally useful for visualization purposes and for gaining insights into the underlying structure of data that might not be apparent in its original high-dimensional form~\cite{borg2005modern,PhysRevD.95.024031,PhysRevResearch.2.043308,Skinner_2021}, especially when there is already some intuition as to which measures are appropriate.

Given $M$ objects which lie in some high-dimensional data space and are defined by pairwise distances $d_{i,j}$, MDS finds a mapping into a chosen $d$-dimensional space while preserving the pairwise distances between the data as much as possible. The output embedded data $x_{1}, ...,x_{M} \in R^{d}$ is such that $||\bm{x}_{i} - \bm{x}_{j}|| \approx d_{i,j}$ for all $i, j \in 1,...,M$. The particular embedding is determined through minimisation of the following loss function
\begin{equation}
    S_{MDS} = \sum_{i<j} (||\bm{x}_{i} - \bm{x}_{j}|| - d_{i,j})^{2},
\end{equation}
which is typically carried about by numerical optimization techniques such as gradient descent or stress majorization \cite{Kruskal_1964, Borg_1997}. Note that the specific algorithms used, as well as differences in initialization and presence of noisy data, result in an embedding that is not necessarily unique.  However, while the specific configurations of points in the lower-dimensional space may vary, the goal of MDS is to capture the essential relationships of the data, and this method is often used in conjunction with other methods to further explore and validate the data structure. 

Generally, considering a set of snapshots of the evolution of a dynamical system, the compressed representation should be isomorphic to the latent underlying intrinsic state of the system. In this work, we employ MDS to reveal potential redundancies in the high-dimensional dynamical full state wavefunction data of the PXP model and demonstrate that the low-dimensional representation obtained via MDS facilitates the identification of scar phases. In order to capture the relationships that we wish to preserve in the embedding, a suitable distance measure between the data must first be chosen.

Here, we consider a probabilistic earth mover's (PEM) distance between two states $\psi_{A}$ and $\psi_{B}$ which is inspired by the similarly-named metric from optimal transport theory \cite{Skinner_2021, De_Palma_2021}. If $P_{A}(i)$ is the probability of obtaining basis state $i$ when the system is in state $\psi_{A}$ and $P_{B}(j)$ the probability of obtaining outcome $j$ when the system is in state $\psi_{B}$, the PEM is defined as
\begin{equation}
    d_{EM}(\psi_{A},\psi_{B}) = \min_{\gamma}\sum_{i,j}\gamma_{ij}d(i,j),
\end{equation}
where $\gamma_{ij} \geq 0$ is a coupling map between the basis states $i$ and $j$ of $\psi_{A}$ and $\psi_{B}$ such that $\sum_{j}\gamma_{ij}=P_{A}(i)$ and $\sum_{i}\gamma_{ij}=P_{B}(j)$. The distance between basis states is given by the Hamming distance $d(i,j)$ and the minimum is taken over all
possible couplings $\gamma = (\gamma_{ij})$.

In order to detect scarring through the state dynamics, an appropriate distance measure needs to reflect the structure of the dynamics and capture its constraints. We emphasize that the definition of PEM uses physically relevant information encoded in the metric structure $d(i,j)$ of the underlying observable space. The dynamics generated by the PXP Hamiltonian involves effective single spin flips and so using the PEM metric based on the Hamming distance captures an effective minimal transition time between two different basis states. To capture the dynamics of the system in the MDS embedding, we take the full state wavefunction $\psi_{m}(t_{n})$ at $n=1,...,N$ timesteps, quenched from $m=1,...,M$ initial states as our input data. A distance matrix for each initial state is obtained by finding the pairwise distances between the states $\psi_{m}(t_{i})$, $\psi_{m}(t_{j})$ under the dynamics-inspired PEM metric. After applying the MDS embedding, the outputs for each evolution from a different initial state are $N$ vectors in a lower-dimensional space. Here, we find that an embedding into only two dimensions is sufficient to uncover the dynamical characteristics of the scar manifold. 

\begin{figure}
  \centering
  {\includegraphics[width=\textwidth]{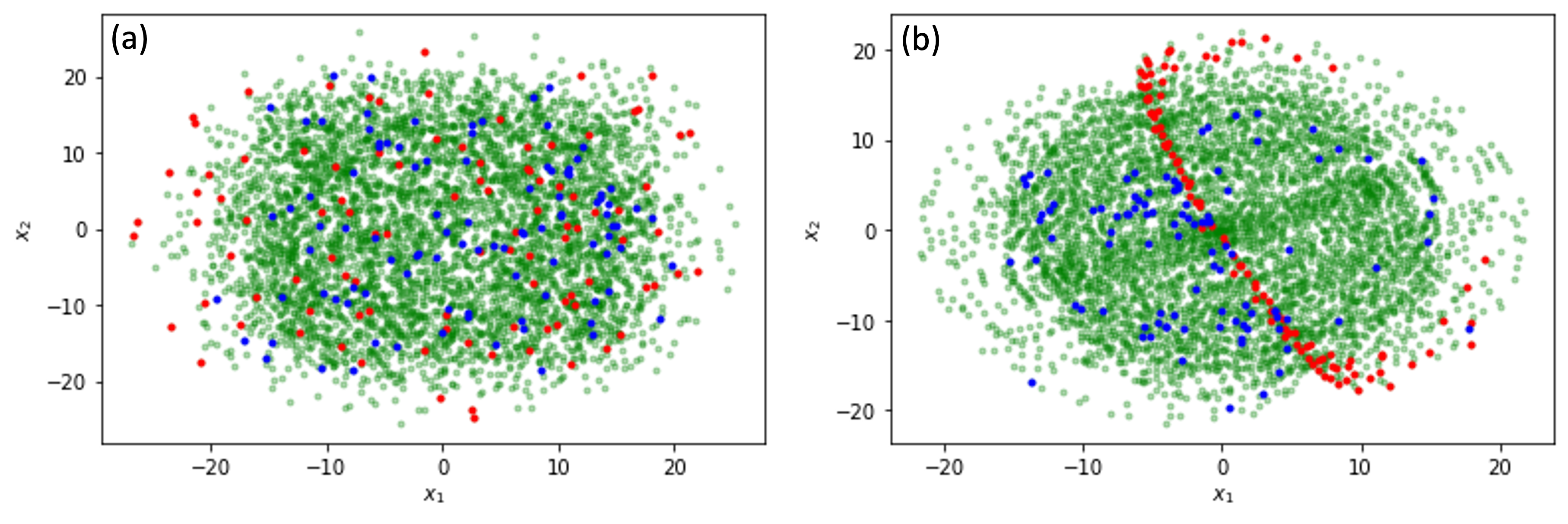}}%
  \caption{Multidimensional scaling 2D embedding of PXP dynamical state (L=10) using (a) Hilbert-Schmidt distance and (b) PEM distance, quenched from all possible initial spin configurations (green). Embedded points for a single evolution from a thermal initial state (blue) and scar initial state (red). Note that the projection axes $(x_{1}, x_{2})$ generally differ for different trajectories and have been aligned for illustration purposes.}
  \label{fig:fig3}
\end{figure}

Figure \ref{fig:fig3}(a) shows that using the Hilbert-Schmidt distance in MDS does not reveal any distinct structural features that allow the thermal and scar initial states to be distinguished in the 2-D embedded space. By contrast, it is evident that using the PEM metric provides considerably more insight into the dynamical structure of the scarring behaviour compared to this standard distance measure, as illustrated by the plot in Fig. \ref{fig:fig3}(b). This tighter arrangement for the embedding of the scar initial state dynamics reflects that, within the PEM metric, the states are bound closer together and require fewer degrees of freedom to describe. Similarly dense structures are observed for other special initial states, such as $\ket{Z_{3}}$.  

In order to appreciate how the PEM metric gives us the ability to distringuish between scar and thermal phases, we notice that it hinges on probability amplitudes of bitstrings rather than state fidelities
\begin{equation}
    |\psi_{n}(t)|^{2} = \sum_{ij}c^{*}_{i}c_{j}\bra{\phi_{i}}\ket{\bm{n}}\bra{\bm{n}}\ket{\phi_{j}}e^{-i(E_{j}-E_{i})t},
\label{ci}
\end{equation}
where $\bm{n}$ is a given bitstring measured in the computational basis, $\phi_{i}$ are the energy eigenstates, $c_{i}$ is the amplitude with which the eigenstate $\phi_{i}$ is excited,  and $E_{i}$ are the energy eigenvalues. 

\begin{figure}
  \centering
  {\includegraphics[width=0.6\textwidth]{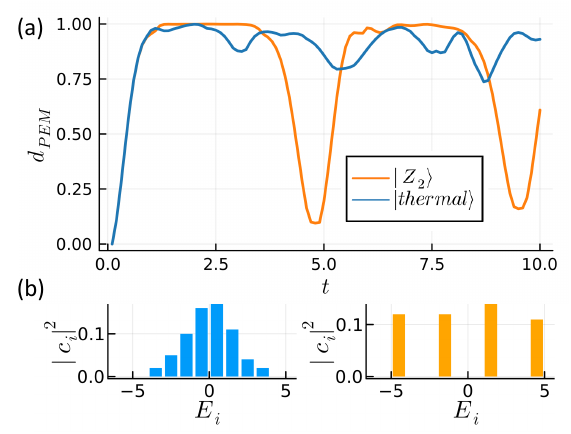}}%
  \caption{Plots showing (a) probabilistic earth mover's metric between initial and evolved state for scar (orange) and thermal (blue) initial states and (b) corresponding distributions of expansion coefficients in energy eigenbasis, sampled over 100 evolutions.}
  \label{fig:fig4}
\end{figure}

This measure is predominantly dependent on two quantities: the projection of the initial state on the computational basis and the distribution of expansion coefficients in the energy eigenbasis. For thermal initial states, the individual initial state amplitudes projected onto the $i$th eigenstate become vanishingly small during the system evolution due to spreading across many states. Furthermore, we find that the expansion coefficients in the energy eigenbasis are distributed randomly (see Fig. \ref{fig:fig4}(b)), which leads to dephasing over terms in the distance measure. 

\section{Intrinsic dimension of scar dynamics}

In a typical experiment, access to measures based on the full wavefunction data is limited because exponentially many measurements are required to define a complete set of state parameters \cite{James_2001}. The following section proposes a potential avenue for overcoming the experimental limitations of the scheme discussed in Section III, by targeting the dimensionality reduction process itself rather than the embedded result. This allows for capturing the desired structural features of the system without a complete description of the dynamics. Specifically, we target a property known as intrinsic dimension (ID), which describes the number of variables required for minimal representation of a dataset without significant information loss \cite{Campadelli_2015, Granata_2016}. In general for natural data sets, the ID is expected to be much lower than the dimensionality of the data due to the presence of structural correlations between variables. This concept can be visualized by imagining a set of Cartesian data points extracted from a circle, as visualised in Fig. \ref{fig:fig1}. Although two input coordinates are required to describe the data, they are collected from a manifold with ID = 1 which causes the points to be strongly correlated. 

Whilst this quantity is often considered implicitly in dimensionality reduction techniques to determine to what extent a dataset can be compressed into a lower-dimension, information about the ID can also be used directly to characterize a system in an unsupervised manner. Recently, ID estimation was used to study phase transitions in spin lattices and dynamics of BECs, showing that ID can be seen as an order parameter which signals a transition between different structures in configuration space ~\cite{PhysRevX.11.011040, PRXQuantum.2.030332, panda_2023, verdel_2023}. For dimensionality reduction methods, the dimension of the identified subspace is often viewed as an estimation of the ID. However, these methods generally come with various conditions which make the estimation of ID a nontrivial task. For example, PCA fails to be effective if the lower dimensional manifold is curved or nonlinear. As a result, various methods have been developed which provide a direct estimate for the ID from the data set as a whole, without any projection steps that may affect the structural relationships within the data \cite{Camastra_2016, Facco_2017}.

The technique used here is explicitly formulated for discrete spaces and works by considering statistics of points within a given volume on the lattice \cite{Macocco_2023}. Since we would like to improve the experimental accessibility of the method, instead of the full state probability distribution we take as data projective measurements of the state at different evolution times. Specifically, we sample $P$ bitstrings from each wavefunction $\psi_{m}(t_{n})$ at $n=1,...,N$ timesteps. This set of bitstrings for each initial state quench $m$ are then respectively input into the ID estimation method with the Hamming distance as a metric. 

\begin{figure}
  \centering
  {\includegraphics[width=0.8\textwidth]{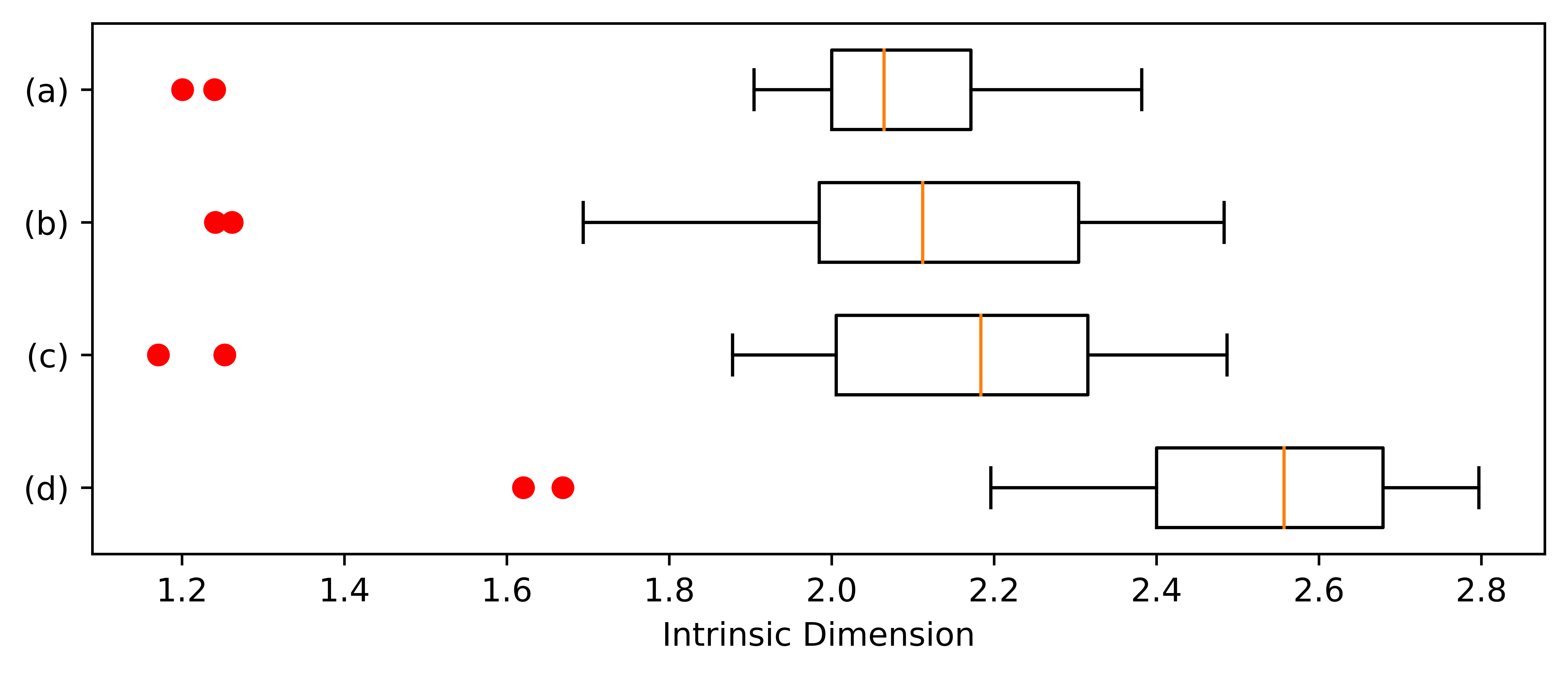}}%
  \caption{Boxplots showing the quartile summaries of the intrinsic dimension estimation for varying shots per time step $N$, time steps $P$ and measurement errors $\epsilon$: (a) $N=500, P=20$, (b) $N=50, P=10$, (c) $N=10, P=50$, (d)$N=50, P=10, \epsilon=0.05$. The orange vertical line corresponds to the median of the data and the red points correspond to fliers lying outside of the inter-quartile range.}
  \label{fig:fig5}
\end{figure}

In Fig. \ref{fig:fig5}, the resulting ID estimates are summarised using boxplots, a statistical representation of the data. The three verticle lines forming the central boxes correspond to the first quartile, median and third quartile of the data respectively. The outermost 'whiskers' extend from the box to the farthest data points lying within 1.5x the inter-quartile range (IQR) from the box. An emphasis is placed on outlying data (fliers) shown in red points, which allows a distinction to be seen between the thermal majority and scarring initial states. Figure \ref{fig:fig5}(a) shows the estimated values from samples of each initial configuration in the large measurement limit. It is evident that there are two initial states whose sample ID is considerably lower than the rest. These configurations correspond respectively to the $\ket{Z_{2}}$ and $\ket{Z_{2}'}=\ket{10\cdot\cdot\cdot10}$ states, which are the special initial states that result in scar dynamics for the PXP model. This is an unsurprising result as we are effectively considering the full probability distributions which we noticed in Section III, when the scar state dynamics are embedded in a lower-dimension, are restricted in the embedding space as compared to the thermal states. The estimated ID value of $\sim$1.2 can be interpreted as the data having approximately an ID of $\sim$1 but containing noise which inflates the dimension slightly. 

The main band of points spread around ID $\approx 2$ correspond to the thermal majority of initial states. This value of the ID similarly agrees with the results from Fig. \ref{fig:fig3}(b), which suggests that when the wavefunction data is projected using MDS, the thermal initial states result in an embedding which extends across a two-dimensional space. These dimensions arise from the time evolution of the dynamical system and the deviation away from the scar manifold. The variation can again be attributed the presence of noise, although we note that, the $\ket{Z_{3}}$ and other periodic charge-density states can be found lying at the bottom of the band. These states have been found to exhibit some characteristics of scarring behaviour such as periodic quantum revival, although with much higher rates of decay \cite{Turner_2018}. 


Here, we investigate the robustness of this ID distinguishability between scar and thermal initial states under experimentally realistic limitations. This includes reducing the number of samples made at each timestep and the total number of timesteps, as well as introducing sampling errors. These are readout errors caused by erroneous measurement operations which lead to sites being incorrectly identified as being in the ground or excited state. We consider error rates up to typical experimental error rates of around 1\% \cite{ionq_2023}.

Figures \ref{fig:fig5}(b) and \ref{fig:fig5}(c) illustrate that with limited measurement times $P$ or limited number of measurements $N$ respectively, the range and IQR are increased but the ID remains sufficiently distinct between thermal and scar dynamics. We note that when both $P$ and $N$ are reduced significantly, the failure of the data to sufficiently represent the dynamics causes this distinction to become blurred out. In a similar fashion, Fig. \ref{fig:fig5}(d) shows the ID estimation boxplot when measurement errors are applied to the shot samples. Up to an error rate of 5\%, the ID estimation can still be used to identify the scar initial states. Note that the introduction of noise shifts the ID estimation of all states upwards due to the addition of bitstrings normally forbidden by the constraints of the PXP model. However, the magnitude of this shift is constant between the two dynamical phases and so the distinguishability is unaffected.

\section{Discussion and Outlook}

Quantum many-body scar models exhibit a deviation from the expected thermal behaviour of many-body systems, resulting in a robustness to decoherence that sees potential applications in quantum control and quantum metrology \cite{Bluvstein_2021, Dooley_2021, Desaules_2022}. Without a priori knowledge of the model and access to its full spectrum or entanglement scaling, it is difficult to determine the initial states for system quenching which allow the properties of scarring to be observed in the dynamics. Thus, we employ techniques from machine learning, and in particular dimensionality reduction, to learn these scar properties directly from experimentally accessible data. A low-dimensional embedding of the full wavefunction evolution using multidimensional scaling reveals structural distinctions between thermal and scar states and is suggestive of dynamics in a lower-dimensional manifold than the state space. We aim to learn structural properties of this manifold directly from the intrinsic dimension of the system dynamics, which we estimate using a recent method designed for discrete spaces. As such, this method can be applied in an unsupervised fashion to shot data sampled from the system at different evolution times when quenched from different initial product state configurations. The results indicate that the intrinsic dimension for scarring and thermalizing states can be used to distinguish between these phases. This method is also shown to be robust to limited samples and measurement times, as well as typical experimental measurement errors. 

The discovery of many-body scarring arose as a result of experimental advancements in cold atom setups using optical tweezers and now there have already been a variety of scarring behaviours observed in other experimental platforms, such as superconducting qubit processors and bosonic quantum gases \cite{Kao_2021, Zhang_2022}. We envisage that the approach introduced in this work provides a general and experimentally accessible method of detecting scarring states, requiring only a (polynomially) small number of measurements in the computational basis and being robust to generic measurement errors. Our work also provides a novel perspective on the analysis of QMBS from the perspective of dynamical manifold intrinsic dimension, which may see potential theoretical links to network and complexity theory in future studies \cite{Bhattacharjee_2022, Vitale_2023, mendessantos_2023, nandy_2023}.

\section*{Acknowledgements}

This research is supported by the National Research Foundation, Singapore and A*STAR under its CQT Bridging Grant and Quantum Engineering Programme NRF2021-QEP2-02-P02, A*STAR (\#21709) and by EU HORIZON- Project 101080085 — QCFD.

\bibliography{references}

\appendix

\section{Intrinsic dimension estimation for discrete spaces}

Here we outline the method used in this work for ID estimation of discrete data from \cite{Skinner_2021}. Assuming a set of independently generated data points with density $\rho$, the probability of observing $n$ points in a given region $A$ follows a Poisson distribution. Now, consider a specific data point $i$ which lies within two regions $A$ and $B$, where one contains the other, such that $i \in A \subset B$. It can be shown that the conditional probability of having $n$ points in $A$ given that there are $k$ points in $B$ follows a binomial distribution whose success probability $p$ is a ratio of the volumes of the two regions. Remarkably, this probability is independent of the density $\rho$, which allows the estimator to be used even when the density varies across large distance scales. 

Now, we can write a likelihood function of the observations $n_{i}$ as
\begin{equation}
    L(n_{i}|k_{i}, p_{i}) = \prod^{N}_{i=1} \binom{k_{i}}{n_{i}}p_{i}^{n_{i}}(1- p_{i})^{k_{i} - n_{i}},
    \label{likelihood}
\end{equation}
where $p_{i}$ and $k_{i}$ are parameters which may possibly be point-dependent. Taking our space to be a lattice with the $L_{1}$ metric as a natural choice, the volume of a given region $V(A)$ is simply the number of lattice points contained in $A$. According to Ehrhart theory of polytopes, the number of lattice points within distance $t$ in dimension $d$ from a given point amounts to 
\begin{equation}
    V(t,d) = {d+t\choose d}{}_{2}F_{1}(-d, -t, -d-t, -1),
    \label{volume}
\end{equation}
where ${}_{2}F_{1}(a,b,c,z)$ is the ordinary hypergeometric function. For a given $t$, this is a polynomial in $d$ of order $t$, which makes the ratio of volumes in the likelihood function a ratio of polynomials in $d$. Given a dataset, the choice of two distances $t_1$ and $t_2$ which define the scale at which the data is probed fixes the likelihood function in Eq. (\ref{likelihood}). This can be maximised with respect to the dimension $d$ to infer an estimate for the ID of the data manifold, given by the root of the following equation which can be obtained via standard optimization techniques
\begin{equation}
    \frac{V(t_{1},d)}{V(t_{2},d)} - \frac{\langle n \rangle}{\langle k \rangle} = 0,
\end{equation}
where the mean values over $n$ and $k$ are computed over all data points \cite{Macocco_2023}. Note that, the observation resolution determined by choices of $t_1$ and $t_2$ has considerable impact on the estimated ID if the scale is comparable to the noise level and curvature of the data. Thus, the ID is first estimated over a range of scales and then decided upon by seeing at which value it plateaus, indicating the scale at which the global ID dominates. 

\end{document}